\documentclass[aps,prd,preprint,showpacs,groupaddress]{revtex4}

\usepackage{graphicx}

\begin{document}

\title{A Compromise between Neutrino Masses and Collider
Signatures in the Type-II Seesaw Model}

\author{\bf Wei Chao}
\email{chaowei@mail.ihep.ac.cn}

\author{\bf Shu Luo}
\email{luoshu@mail.ihep.ac.cn}

\author{\bf Zhi-zhong Xing}
\email{xingzz@mail.ihep.ac.cn}

\author{\bf Shun Zhou}
\email{zhoush@mail.ihep.ac.cn}

\affiliation{Institute of High Energy Physics, Chinese Academy of
Sciences, P.O. Box 918, Beijing 100049, China}

\begin{abstract}
A natural extension of the standard $SU(2)^{}_{\rm L} \times
U(1)^{}_{\rm Y}$ gauge model to accommodate massive neutrinos is
to introduce one Higgs triplet and three right-handed Majorana
neutrinos, leading to a $6\times 6$ neutrino mass matrix which
contains three $3\times 3$ sub-matrices $M^{}_{\rm L}$, $M^{}_{\rm
D}$ and $M^{}_{\rm R}$. We show that three light Majorana
neutrinos (i.e., the mass eigenstates of $\nu^{}_e$, $\nu^{}_\mu$
and $\nu^{}_\tau$) are exactly massless in this model, if and only
if $M^{}_{\rm L} = M^{}_{\rm D} M^{-1}_{\rm R} M^T_{\rm D}$
exactly holds. This no-go theorem implies that small but
non-vanishing neutrino masses may result from a significant but
incomplete cancellation between $M^{}_{\rm L}$ and $M^{}_{\rm D}
M^{-1}_{\rm R} M^T_{\rm D}$ terms in the Type-II seesaw formula,
provided three right-handed Majorana neutrinos are of ${\cal
O}(1)$ TeV and experimentally detectable at the LHC. We propose
three simple Type-II seesaw scenarios with the $A^{}_4 \times
U(1)^{}_{\rm X}$ flavor symmetry to interpret the observed
neutrino mass spectrum and neutrino mixing pattern. Such a
TeV-scale neutrino model can be tested in two complementary ways:
(1) searching for possible collider signatures of lepton number
violation induced by the right-handed Majorana neutrinos and
doubly-charged Higgs particles; and (2) searching for possible
consequences of unitarity violation of the $3\times 3$ neutrino
mixing matrix in the future long-baseline neutrino oscillation
experiments.
\end{abstract}


\maketitle

\section{Introduction}
The solar \cite{SNO}, atmospheric \cite{SK}, reactor \cite{KM} and
accelerator \cite{K2K} neutrino experiments have provided us with
very convincing evidence that neutrinos are massive and lepton
flavors are mixed. This important discovery indicates that the
Standard Model (SM), in which neutrinos are massless and lepton
flavors are conserved, is actually incomplete. In order to
generate tiny neutrino masses, one may naturally extend the SM by
introducing three right-handed Majorana neutrinos and one Higgs
triplet but preserving the $SU(2)_{\rm L} \times U(1)_{\rm Y}$
gauge symmetry. The relevant Lagrangian for lepton masses can be
written as
\begin{eqnarray}
- {\cal L}^{}_{\rm lepton} = \overline{l^{}_{\rm L}} Y^{}_l
\tilde{H} E^{}_{\rm R} + \overline{l^{}_{\rm L}} Y^{}_{\nu}
N^{}_{\rm R} H + \frac{1}{2} \overline{N^c_{\rm R}} M^{}_{\rm R}
N^{}_{\rm R} + \frac{1}{2} Y^{}_\Delta \overline{l^{}_{\rm L}}
i\sigma^{}_2 \Delta^{}_{\rm L} l^c_{\rm L} + {\rm h.c.} \; ,
\end{eqnarray}
where $l^{}_{\rm L}$ is the lepton doublet, $H$ with $\tilde{H}
\equiv i\sigma^{}_2 H^*$ is the Higgs doublet, $E^{}_{\rm R}$ and
$N^{}_{\rm R}$ stand respectively for the $SU(2)^{}_{\rm L}$
singlets of charged leptons and neutrinos, and $\Delta^{}_{\rm L}$
denotes the Higgs triplet. After spontaneous symmetry breaking, we
obtain the mass matrices $M^{}_l = Y^{}_l v$, $M^{}_{\rm L} =
Y^{}_\Delta v^{}_{\rm L}$ and $M^{}_{\rm D} = Y^{}_\nu v$, where
$v = \langle H^0 \rangle $ and $v^{}_{\rm L} = \langle
\Delta^{}_{\rm L} \rangle$ are the vacuum expectation values
(vev's) of the neutral components of scalar fields $H$ and
$\Delta^{}_{\rm L}$, respectively. A precision measurement of the
$\rho$-parameter \cite{PDG} strictly constrains the tree-level
contribution of the Higgs triplet to the SM, and thus we are left
with $v^{}_{\rm L} \lesssim 1~{\rm GeV}$ together with $v \approx
174~{\rm GeV}$. The mass scale of $M^{}_{\rm R}$, which is not
subject to the gauge symmetry breaking scale, can be much higher
than $v$. To the leading order, the effective mass matrix for
three light neutrinos is determined by the Type-II seesaw formula
$M^{}_\nu \approx M^{}_{\rm L} - M^{}_{\rm D} M^{-1}_{\rm R}
M^T_{\rm D}$ \cite{SS,typeII}. If the Higgs triplet
$\Delta^{}_{\rm L}$ is absent, the small mass scale of $M^{}_\nu$
can be just attributed to the large mass scale of $M^{}_{\rm R}$
(i.e., the Type-I seesaw mechanism \cite{SS}). In the absence of
heavy right-handed Majorana neutrinos, the observed smallness of
three neutrino masses implies that the mass scale of $M^{}_{\rm
L}$ should be extremely small. A general case is that both terms
of $M^{}_\nu$ are important (e.g., comparable in magnitude) and
their significant cancellation leads to small neutrino masses. In
connection with the origin of neutrino masses, the phenomenon of
lepton flavor mixing arises from the mismatch between the
diagonalizations of $M^{}_l$ and $M^{}_\nu$.

Seesaw mechanisms are currently the most natural way to generate
tiny neutrino masses, and they can naturally be embedded into more
fundamental frameworks such as the grand unified theories (GUT's)
or string theory. Typical examples of this nature are the $SO(10)$
GUT's \cite{GUT} and the $E_8 \times E_8$ heterotic string theory
\cite{string}. A salient feature of most seesaw models is that the
thermal leptogenesis mechanism \cite{FY} can work well to account
for the cosmological baryon number asymmetry via the CP-violating
and out-of-equilibrium decays of heavy right-handed neutrinos and
the $(B-L)$-conserving sphaleron processes. On the experimental
side, however, how to test seesaw mechanisms has been a question.
Given the light neutrino mass scale $m^{}_\nu \sim 0.01~{\rm eV}$
and $Y^{}_\nu \sim {\cal O}(1)$ in the Type-I seesaw scenario, the
mass scale of right-handed Majorana neutrinos is expected to be
$m^{}_{\rm R} \sim 10^{15}~{\rm GeV}$ as a straightforward
consequence of the inverted seesaw formula $M^{}_{\rm R} \approx -
M^{T}_{\rm D} M^{-1}_\nu M^{}_{\rm D}$. Such neutral particles can
never be produced and detected at any colliders even in the far
future, not only because they are too heavy but also because the
strength of their charged-current interactions (characterized by
the ratio $M^{}_{\rm D}M^{-1}_{\rm R} \sim
\sqrt{m^{}_\nu/m^{}_{\rm R}} \sim 10^{-13}$) is too small. A
possible way out is to lower the mass scale of $M^{}_{\rm R}$ down
to the TeV level but allow the Yukawa coupling matrix $Y^{}_\nu$
to be of ${\cal O}(10^{-3})$ up to ${\cal O}(1)$. In order to
generate sufficiently small neutrino masses in this kind of
TeV-scale seesaw scenarios \cite{Pilaftsis,early}, the key point
is to adjust the textures of $M^{}_{\rm D}$ and $M^{}_{\rm R}$ to
guarantee $M^{}_{\rm D} M^{-1}_{\rm R} M^T_{\rm D} = 0$ in the
leading-order approximation. Then tiny but non-vanishing neutrino
masses can be ascribed to slight perturbations or radiative
corrections to $M^{}_{\rm D} M^{-1}_{\rm R} M^T_{\rm D}$ in the
next-to-leading order approximation. Although such a seesaw model
seems quite contrived, it is hopeful to be tested at the Large
Hadron Collider (LHC) by searching for clear
lepton-number-violating signals induced by heavy Majorana
neutrinos \cite{Han}. Recently, Kersten and Smirnov \cite{smirnov}
have reconsidered this sort of {\it structural cancellation} in
the Type-I seesaw formula and pointed out some possible flavor
symmetries behind it. One of their important observations is that
the main structures of $M^{}_{\rm D}$ and $M^{}_{\rm R}$, which
are relevant to possibly observable collider signatures, are
difficult to imprint on those sub-leading effects (due to explicit
perturbations or radiative corrections) responsible for tiny
neutrino masses. In other words, collider physics seems to be
essentially decoupled from neutrino physics in generic Type-I
seesaw scenarios \cite{smirnov}, no matter whether the heavy
Majorana neutrinos are of ${\cal O}(1)$ TeV or much heavier than
that.

This work aims to extend Kersten and Smirnov's consideration to
the Type-II seesaw case with both the right-handed Majorana
neutrinos and the Higgs triplet at the TeV scale. This extension
is non-trivial and intriguing at least in two aspects: (a) instead
of realizing the structural cancellation (i.e., $M^{}_{\rm D}
M^{-1}_{\rm R} M^T_{\rm D} \approx 0$), we consider the {\it
global cancellation} between the contribution from $\Delta^{}_{\rm
L}$ and that from right-handed Majorana neutrinos (i.e.,
$M^{}_{\rm L} - M^{}_{\rm D} M^{-1}_{\rm R} M^T_{\rm D} \approx
0$); (b) not only the TeV-scale Majorana neutrinos but also the
doubly-charged components of $\Delta^{}_{\rm L}$ are possible to
show up in the collider experiments. In fact, the long-lived
doubly-charged scalar has already been searched for at the
Tevatron \cite{D0}. We shall prove a no-go theorem: the masses of
light Majorana neutrinos are exactly vanishing at the tree level
if and only if the global cancellation $M^{}_{\rm L} - M^{}_{\rm
D} M^{-1}_{\rm R} M^T_{\rm D} = 0$ exactly holds in generic
Type-II seesaw scenarios. Therefore, a feasible way to obtain both
tiny neutrino masses and appreciable collider signatures is to
allow for an incomplete cancellation between $M^{}_{\rm L}$ and
$M^{}_{\rm D} M^{-1}_{\rm R} M^T_{\rm D}$ terms. To be explicit,
we shall propose three simple type-II seesaw scenarios with the
$A^{}_4 \times U(1)^{}_{\rm X}$ flavor symmetry at the TeV scale,
from which the observed neutrino mass spectrum and neutrino mixing
pattern can be achieved. We shall also discuss two interesting
consequences of this model: (1) possible unitarity violation of
the $3\times 3$ neutrino mixing matrix, which can be searched for
in the future long-baseline neutrino oscillation experiments; and
(2) possible signatures of lepton number violation induced by the
right-handed Majorana neutrinos and doubly-charged Higgs
particles, which can be searched for at the LHC and other
colliders.

The remaining part of this paper is organized as follows. In
section II, we review some basics of the type-II seesaw mechanism
and prove the no-go theorem. Section III is devoted to a specific
type-II seesaw model, in which the incomplete cancellation between
$M^{}_{\rm L}$ and $M^{}_{\rm D} M^{-1}_{\rm R} M^T_{\rm D}$ terms
is realized by the $A^{}_4 \times U(1)^{}_{\rm X}$ symmetry and
its breaking. The unitarity violation of the $3\times 3$ neutrino
mixing matrix and possible collider signatures of lepton number
violation are discussed in section IV. Some conclusions are drawn
in section V.

\section{Type-II Seesaw and No-go Theorem}

We regularize our notations and conventions in this section by
reviewing some basics of the Type-II seesaw mechanism. After
spontaneous symmetry breaking, the lepton mass terms in Eq. (1)
turn out to be
\begin{eqnarray}
-{\cal L}_{\rm mass} = \overline{E^{}_{\rm L}} M^{}_l E^{}_{\rm R} +
\frac{1}{2} \overline{\left( \nu^{}_{\rm L} ~N^c_{\rm R}\right)}
\left( \matrix{ M^{}_{\rm L} & M^{}_{\rm D} \cr M^T_{\rm D} &
M^{}_{\rm R}}\right) \left( \matrix{ \nu^c_{\rm L} \cr N^{}_{\rm
R}}\right) + {\rm h.c.} \; ,
\end{eqnarray}
where $\nu^c_{\rm L} \equiv C \overline{\nu^{}_{\rm L}}^T$ with
$C$ being the charge conjugation matrix, likewise for $N^c_{\rm
R}$. The overall $6\times 6$ neutrino mass matrix in ${\cal
L}^{}_{\rm mass}$, denoted as ${\cal M}$, can be diagonalized by
the unitary transformation ${\cal U}^\dagger {\cal M} {\cal U}^* =
\widehat{\cal M}$; or explicitly,
\begin{eqnarray}
\left(\matrix{V & R \cr S & U}\right)^\dagger \left( \matrix{
M^{}_{\rm L} & M^{}_{\rm D} \cr M^T_{\rm D} & M^{}_{\rm R}}\right)
\left(\matrix{V & R \cr S & U}\right)^*  = \left( \matrix{
\widehat{M}^{}_\nu & {\bf 0} \cr {\bf 0} & \widehat{M}^{}_{\rm
N}}\right) \; ,
\end{eqnarray}
where $\widehat{M}^{}_\nu = {\rm Diag}\{m^{}_1, m^{}_2, m^{}_3\}$
and $\widehat{M}^{}_{\rm N} = {\rm Diag}\{M^{}_1, M^{}_2,
M^{}_3\}$ with $m^{}_i$ and $M^{}_i$ (for $i=1, 2, 3$) being the
light and heavy Majorana neutrino masses, respectively. Note that
the $3\times 3$ rotation matrices $V$, $U$, $R$ and $S$ are
non-unitary, but they are correlated with one another due to the
unitarity of $\cal U$:
$$
\begin{array}{rcl}
V^\dagger V + S^\dagger S = V V^\dagger + R R^\dagger = {\bf 1} \; ,
\\
U^\dagger U + R^\dagger R = U U^\dagger + S S^\dagger = {\bf 1} \; ,
\end{array}
\eqno{\rm (4a)}
$$
and
$$
R^\dagger V + U^\dagger S = S V^\dagger + U R^\dagger = {\bf 0} \; .
\eqno{\rm (4b)}
$$
The effective neutrino mass matrix $M^{}_\nu$ can be defined by
decomposing ${\cal U}$ into a product of two unitary matrices
${\cal W}$ and $\cal V$:
\setcounter{equation}{4}
\begin{eqnarray}
{\cal V}^\dagger {\cal W}^\dagger \left( \matrix{ M^{}_{\rm L} &
M^{}_{\rm D} \cr M^T_{\rm D} & M^{}_{\rm R}}\right) {\cal W}^* {\cal
V}^* \equiv {\cal V}^\dagger \left( \matrix{ M^{}_\nu & {\bf 0} \cr
{\bf 0} & M^{}_{\rm N}}\right) {\cal V}^* = \left( \matrix{
\widehat{M}^{}_\nu & {\bf 0} \cr {\bf 0} & \widehat{M}^{}_{\rm
N}}\right) \; ,
\end{eqnarray}
where ${\cal W}$ and ${\cal V}$ take the general forms
\begin{eqnarray}
{\cal W} = \left( \matrix{ U^{}_1 & B \cr C & U^{}_2}\right) \; ,
~~~ {\cal V} = \left( \matrix{ V^{}_1 & {\bf 0} \cr {\bf 0} &
V^{}_2}\right) \; .
\end{eqnarray}
The $3\times 3$ rotation matrices $U^{}_1$, $B$, $C$ and $U^{}_2$
are non-unitary, but they satisfy the normalization and
orthogonality conditions of ${\cal W}$ just like the correlative
conditions of $V$, $R$, $S$ and $U$ given in Eq. (4). In contrast,
$V^{}_1$ and $V^{}_2$ are unitary. It is trivial to obtain the
relations $V = U^{}_1 V^{}_1$ and $R = B V^{}_2$ from ${\cal U} =
{\cal W} {\cal V}$. To express $M^{}_\nu$ as a recursive expansion
in powers of $M^{}_{\rm D} M^{-1}_{\rm R}$, an ansatz has been
made for ${\cal W}$ in Ref. \cite{Grimus}, in which
$C=-B^\dagger$, $U^{}_1 = \sqrt{{\bf 1} -B B^\dagger}$ and $U^{}_2
= \sqrt{{\bf 1} -B^\dagger B}$ are reasonably assumed. More
general but less instructive expressions of $M^{}_\nu$ and
$M^{}_{\rm N}$ can be found in Ref. \cite{xingzhou}. To the
leading order,
\begin{eqnarray}
M^{}_\nu \approx M^{}_{\rm L} - M^{}_{\rm D} M^{-1}_{\rm R}
M^T_{\rm D} \; ,
\end{eqnarray}
known as the Type-II seesaw formula.

After diagonalizing ${\cal M}$, one may express the neutrino
flavor eigenstates $\nu^{}_\alpha$ (for $\alpha = e, \mu, \tau$)
in terms of the light and heavy neutrino mass eigenstates
$\nu^{}_i$ and $N^{}_i$ (for $i=1, 2, 3$):
\setcounter{equation}{7}
\begin{eqnarray}
\left(\matrix{ \nu^{}_e \cr \nu^{}_\mu \cr \nu^{}_\tau
}\right)^{}_{\rm L} = V \left( \matrix{\nu^{}_1 \cr \nu^{}_2 \cr
\nu^{}_3} \right)^{}_{\rm L} + R \left( \matrix{N^{}_1 \cr N^{}_2
\cr N^{}_3} \right)^{}_{\rm L} \; .
\end{eqnarray}
In the basis where the flavor eigenstates of three charged leptons
are identified with their mass eigenstates, the standard
charged-current interactions between $\nu^{}_\alpha$ and $\alpha$
(for $\alpha = e, \mu, \tau$) turn out to be
\begin{eqnarray}
-{\cal L}^{}_{\rm cc} = \frac{g}{\sqrt{2}} \left[
\overline{\left(e~~ \mu~~ \tau\right)^{}_{\rm L}} V \gamma^\mu
\left( \matrix{\nu^{}_1 \cr \nu^{}_2 \cr \nu^{}_3} \right)^{}_{\rm
L} W^-_{\mu} + \overline{\left(e~~ \mu~~ \tau\right)^{}_{\rm L}} R
\gamma^\mu \left( \matrix{N^{}_1 \cr N^{}_2 \cr N^{}_3}
\right)^{}_{\rm L} W^-_\mu \right] + {\rm h.c.} \; .
\end{eqnarray}
It becomes clear that $V$ describes the charged-current interactions
of three light Majorana neutrinos $(\nu^{}_1, \nu^{}_2, \nu^{}_3)$,
while $R$ is relevant to the charged-current interactions of three
heavy Majorana neutrinos $(N^{}_1, N^{}_2, N^{}_3)$. One may
similarly write out the interactions between the Majorana neutrinos
and the neutral gauge boson (or Higgs) in the chosen flavor basis
\cite{Pilaftsis}. It is mainly the strength of charged-current
interactions that determines the production and detection
probabilities of heavy Majorana neutrinos at hadron or $e^+e^-$
colliders. To experimentally test a seesaw mechanism, two
prerequisites have to be satisfied: the mass scale of $N^{}_i$
should be low enough and the magnitude of $R$ should be large
enough. But both of them may in general give rise to unacceptably
sizable masses of $\nu^{}_i$ through the seesaw formula. One
possible way to get around this difficulty in the Type-II seesaw
mechanism might be to dictate a complete cancellation between the
leading terms $M^{}_{\rm L}$ and $M^{}_{\rm D} M^{-1}_{\rm R}
M^{T}_{\rm D}$ and generate tiny neutrino masses via the sub-leading
terms of $M^{}_\nu$ in Eq. (7). Such an idea is seemingly
reasonable, but it does not work because of the following no-go
theorem:

{\it If and only if the relationship $M^{}_{\rm L} = M^{}_{\rm D}
M^{-1}_{\rm R} M^T_{\rm D}$ is exactly satisfied in generic
Type-II seesaw models, then three light Majorana neutrinos must be
exactly massless.}

\noindent In other words, imposing the pre-condition $M^{}_{\rm L}
= M^{}_{\rm D} M^{-1}_{\rm R} M^T_{\rm D}$ on the $6\times 6$
neutrino mass matrix ${\cal M}$ will automatically guarantee
$M^{}_\nu = \widehat{M}^{}_\nu = {\bf 0}$ for three light Majorana
neutrinos. Hence tiny neutrino masses can only be generated from
an incomplete cancellation between $M^{}_{\rm L}$ and $M^{}_{\rm
D} M^{-1}_{\rm R} M^T_{\rm D}$ terms or from radiative
corrections. A similar theorem is valid for the canonical seesaw
mechanism by setting $M^{}_{\rm L} =0$; i.e., three light Majorana
neutrinos must be massless if and only if $M^{}_{\rm D}
M^{-1}_{\rm R} M^T_{\rm D} =0$ exactly holds in generic Type-I
seesaw models.

Now let us prove the above theorem in a way without loss of any
generality. Rewriting Eq. (3) as ${\cal M} {\cal U}^* = {\cal U}
\widehat{\cal M}$ and doing the matrix multiplication on both
left- and right-hand sides, we obtain
$$
V \widehat{M}^{}_\nu = M^{}_{\rm L} V^* + M^{}_{\rm D} S^* \; , \\
\eqno{\rm (10a)}
$$
$$
S \widehat{M}^{}_\nu = M^T_{\rm D} V^* + M^{}_{\rm R} S^* \; , \\
\eqno{\rm (10b)}
$$
$$
R \widehat{M}^{}_{\rm N} = M^{}_{\rm L} R^* + M^{}_{\rm D} U^* \; , \\
\eqno{\rm (10c)}
$$
$$
U \widehat{M}^{}_{\rm N} = M^T_{\rm D} R^* + M^{}_{\rm R} U^* \; . \\
\eqno{\rm (10d)}
$$
The first step of our proof is to derive $\widehat{M}^{}_\nu = 0$
from the pre-condition $M^{}_{\rm L} = M^{}_{\rm D} M^{-1}_{\rm R}
M^T_{\rm D}$. Multiplying Eq. (10b) by $M^{}_{\rm D} M^{-1}_{\rm
R}$ on the left and taking account of Eq. (10a) and $M^{}_{\rm L}
= M^{}_{\rm D} M^{-1}_{\rm R} M^T_{\rm D}$, we get
\setcounter{equation}{10}
\begin{eqnarray}
\left(M^{}_{\rm D}M^{-1}_{\rm R}S - V\right) \widehat{M}^{}_\nu =
{\bf 0} \; .
\end{eqnarray}
Multiplying Eq. (10d) by $M^{}_{\rm D} M^{-1}_{\rm R}$ on the left
and taking account of Eq. (10c), we analogously arrive at
\begin{eqnarray}
\left(M^{}_{\rm D}M^{-1}_{\rm R}U - R\right) \widehat{M}^{}_{\rm N}
= {\bf 0} \; .
\end{eqnarray}
By definition, $\widehat{M}^{}_{\rm N}$ is a diagonal matrix
containing three real and positive eigenvalues (i.e., the masses
of three heavy Majorana neutrinos). Hence the unique solution to
Eq. (12) is $R = M^{}_{\rm D} M^{-1}_{\rm R} U$. This result,
together with $SV^\dagger + UR^\dagger =0$ given in Eq. (4b),
leads to
\begin{eqnarray}
M^{}_{\rm D}M^{-1}_{\rm R} SV^\dagger + R R^\dagger = {\bf 0} \; .
\end{eqnarray}
Combining Eqs. (4a) and (13), we are then left with
\begin{eqnarray}
\left(M^{}_{\rm D}M^{-1}_{\rm R}S - V\right) V^\dagger = -{\bf 1}
\; .
\end{eqnarray}
The unit matrix on the right-hand side of Eq. (14) implies that
the ranks of $\left(M^{}_{\rm D}M^{-1}_{\rm R}S - V\right)$ and
$V^\dagger$ must be three, and thus the rank of
$\widehat{M}^{}_\nu$ must be zero as required by Eq. (11). Namely,
$\widehat{M}^{}_\nu =0$ is an unavoidable consequence of
$M^{}_{\rm L} = M^{}_{\rm D} M^{-1}_{\rm R} M^T_{\rm D}$. The
second step of our proof is to show that $M^{}_{\rm L} = M^{}_{\rm
D} M^{-1}_{\rm R} M^T_{\rm D}$ will hold if three light Majorana
neutrinos are massless (i.e., $\widehat{M}^{}_\nu = 0$). For this
purpose, we rewrite Eq. (3) as ${\cal M} = {\cal U} \widehat{\cal
M} {\cal U}^T$ and then impose $\widehat{M}^{}_\nu = 0$ on it.
Three sub-matrices of ${\cal M}$ turn out to be
\begin{eqnarray}
M^{}_{\rm L} = R \widehat{M}^{}_{\rm N} R^T \; ,~~ M^{}_{\rm R} =
U \widehat{M}^{}_{\rm N} U^T \; , ~~ M^{}_{\rm D} = R
\widehat{M}^{}_{\rm N} U^T \; .
\end{eqnarray}
It is easy to verify that $M^{}_{\rm L} = M^{}_{\rm D} M^{-1}_{\rm
R} M^T_{\rm D}$ holds in consequence of Eq. (15), or equivalently
in consequence of $\widehat{M}^{}_\nu = 0$. This completes the
proof of our theorem.

The no-go theorem tells us that it is impossible to generate tiny
neutrino masses from the sub-leading seesaw terms in a recursive
expansion of $M^{}_\nu$ (in powers of $M^{}_{\rm D} M^{-1}_{\rm
R}$), if and only if the condition $M^{}_{\rm L} = M^{}_{\rm D}
M^{-1}_{\rm R} M^T_{\rm D}$ is imposed. This point has more or
less been observed or illustrated in the literature (see, e.g.,
Refs. \cite{Pilaftsis,smirnov,Grimus,Korner}), but only our
present work provides the most general proof without any special
assumption or approximation. In order to reach a compromise
between tiny neutrino masses and accessible collider signatures at
the TeV scale, a phenomenologically viable way is to consider
significant but incomplete cancellation between $M^{}_{\rm L}$ and
$M^{}_{\rm D} M^{-1}_{\rm R} M^T_{\rm D}$ terms in the Type-II
seesaw formula. We shall propose a specific model with the $A^{}_4
\times U(1)^{}_{\rm X}$ flavor symmetry to realize the desired
cancellation in section III and discuss its consequences on
collider physics in section IV.

\section{A specific model with $A^{}_4 \times U(1)^{}_{\rm X}$ symmetry}

To simultaneously achieve tiny neutrino masses and large neutrino
mixing angles, we impose the $A^{}_4 \times U(1)^{}_{\rm X}$
flavor symmetry \cite{A4} on the Type-II seesaw Lagrangian in Eq.
(1). In this case, the assignments of relevant lepton and scalar
fields with respect to the symmetry group $SU(2)^{}_{\rm L} \times
U(1)^{}_{\rm Y} \otimes A^{}_4 \times U(1)^{}_{\rm X}$ are
\begin{eqnarray}
&& l^{}_{\rm L} \sim (2, -1) \otimes (\underline{3}, 1) \; ,
~~~~~~~~~
\phi \sim (2, -1) \otimes (\underline{1}, 1) \; , \nonumber \\
&& E^{}_{\rm R} \sim (1, -2) \otimes (\underline{1}, 1) \; ,
\;\;\; ~~~~~
\Phi \sim (2, -1) \otimes (\underline{3}, 0) \; , \nonumber \\
&& E^\prime_{\rm R} \sim (1, -2) \otimes (\underline{1}^\prime, 1)
\; , ~~~~~~~ \chi \sim (1, 0) \otimes (\underline{3}, 1) \; , \nonumber \\
&& E^{\prime \prime}_{\rm R} \sim (1, -2) \otimes
(\underline{1}^{\prime \prime}, 1) \; , ~~~~~~ \Delta \sim (3, -2)
\otimes (\underline{1}, 2) \; , \nonumber \\
&& N^{}_{\rm R} \sim (1, 0) \otimes (\underline{3}, 0) \; ,
~~~~~~~~~~ \Sigma \sim (3, -2) \otimes (\underline{3}, 0) \; ,
\end{eqnarray}
where several triplet scalars have been introduced. The
irreducible representations of $A^{}_4$ group and the
decomposition of their direct products can be found in Ref.
\cite{He}. Given $SU(2)^{}_{\rm L} \times U(1)^{}_{\rm Y} \otimes
A^{}_4 \times U(1)^{}_{\rm X}$ invariance, the Lagrangian
responsible for lepton masses reads
\begin{eqnarray}
- {\cal L}^{}_{\rm lepton} &=& y^{}_e \left(\overline{l^{}_{\rm
L}} \tilde{\Phi}\right)^{}_{\underline{1}} E^{}_{\rm R} +
y^\prime_e \left(\overline{l^{}_{\rm L}}
\tilde{\Phi}\right)^{}_{\underline{1}^\prime} E^{\prime
\prime}_{\rm R} + y^{\prime \prime}_e \left(\overline{l^{}_{\rm
L}} \tilde{\Phi}\right)^{}_{\underline{1}^{\prime
\prime}} E^{\prime}_{\rm R} \nonumber \\
&& + \frac{1}{2} y^{}_\Delta \overline{l^{}_{\rm L}} i\sigma^{}_2
\Delta l^c_{\rm L} + \frac{1}{2} m^{}_{\rm R}
\left(\overline{N^c_{\rm R}} N^{}_{\rm R}\right)^{}_{\underline{1}}
+ y^{}_\nu \left(\overline{l^{}_{\rm L}}N^{}_{\rm
R}\right)^{}_{\underline{1}} \phi + {\rm h.c.} \; ,
\end{eqnarray}
in which the gauge-invariant and $A^{}_4$-invariant terms
$\overline{l^{}_{\rm L}} N^{}_{\rm R} \Phi$, $\overline{N^c_{\rm
R}} N^{}_{\rm R} \chi$ and $\overline{l^{}_{\rm L}} i\sigma^{}_2
\Sigma l^c_{\rm L}$ do not appear because they are forbidden by
the $U(1)^{}_{\rm X}$ symmetry. After spontaneous symmetry
breaking, the overall neutrino mass matrix $\cal M$ is determined
by its three $3\times 3$ sub-matrices
\begin{eqnarray}
M^{}_{\rm L} = m^{}_{\rm L} \cdot {\bf 1} \; , ~~~~ M^{}_{\rm D} =
m^{}_{\rm D}\cdot {\bf 1} \; , ~~~~ M^{}_{\rm R} = m^{}_{\rm
R}\cdot {\bf 1} \; ,
\end{eqnarray}
where $m^{}_{\rm L} = y^{}_\Delta \langle \Delta \rangle$ and
$m^{}_{\rm D} = y^{}_\nu \langle \phi \rangle$. In the assumption
of $\langle \Phi^{}_1 \rangle = \langle \Phi^{}_2 \rangle =
\langle \Phi^{}_3 \rangle$, the charged-lepton mass matrix can be
written as $M^{}_l = U^{}_l \widehat{M}^{}_l$, where
$\widehat{M}^{}_l = {\rm Diag} \{m^{}_e, m^{}_\mu, m^{}_\tau \} =
\sqrt{3} \langle \Phi^{}_i \rangle {\rm Diag}\{y^{}_e,
y^{\prime}_e, y^{\prime \prime}_e \}$ and
\begin{eqnarray}
U^{}_l = \frac{1}{\sqrt{3}} \left( \matrix{ 1 & 1 & 1 \cr 1 &
\omega & \omega^2 \cr 1 & \omega^2 & \omega}\right) \;
\end{eqnarray}
with $\omega = \exp(i2\pi/3)$. It is quite obvious that $m^{}_{\rm
L} = m^2_{\rm D}/m^{}_{\rm R}$ will lead to $M^{}_{\rm L} =
M^{}_{\rm D} M^{-1}_{\rm R} M^T_{\rm D}$. According to the no-go
theorem, this complete cancellation makes light neutrino masses
exactly vanishing. In order to obtain the realistic neutrino mass
spectrum and lepton flavor mixing pattern, we may introduce an
incomplete cancellation between $M^{}_{\rm L}$ and $M^{}_{\rm D}
M^{-1}_{\rm R} M^T_{\rm D}$ terms by breaking the flavor symmetry
$U(1)^{}_{\rm X}$ explicitly to $Z^{}_2$. The $U(1)^{}_{\rm
X}$-violating terms, such as $(\Phi^\dagger
\phi)^{}_{\underline{3}} \cdot (\Phi^\dagger
\phi)^{}_{\underline{3}}$ in the scalar potential \cite{He}, can
accomplish this purpose. For simplicity, we list the complete
scalar potential in Appendix A. Note that the explicit breaking of
the global $U(1)^{}_{\rm X}$ symmetry does not yield the
problematic Goldstone particle. We may assign the proper $Z^{}_2$
parity to produce slight perturbations to the neutrino mass terms.
Three possibilities are discussed in order.

(1) Perturbations to $M^{}_{\rm L}$: $l^{}_{\rm L}$, $E^{}_{\rm
R}$, $E^{\prime}_{\rm R}$, $E^{\prime \prime}_{\rm R}$, $\chi$ and
$\phi$ are odd under the $Z^{}_2$ transformation, while the other
fields are even under the same transformation. In this case, the
Yukawa interaction $y^{}_{\Sigma} \overline{l^{}_{\rm L}}
i\sigma^{}_2 \Sigma l^c_{\rm L}$ is no longer forbidden and it
contributes a few off-diagonal terms to the effective neutrino
mass matrix:
\begin{eqnarray}
M^{}_\nu = \delta m \cdot {\bf 1} + \left(\matrix{ {\bf 0} &
\omega^{}_3 & \omega^{}_2 \cr \omega^{}_3 & {\bf 0} & \omega^{}_1
\cr \omega^{}_2 & \omega^{}_1 & {\bf 0}} \right) \; ,
\end{eqnarray}
where $\delta m = m^{}_{\rm L} - m^2_{\rm D}/m^{}_{\rm R}$ is the
residue of the incomplete cancellation induced by the mass terms
in Eq. (18), and $\omega^{}_i =y^{}_\Sigma \langle \Sigma^{}_i
\rangle$ (for $i=1, 2, 3$). In the assumption of $\langle
\Sigma^{}_1 \rangle = \langle \Sigma^{}_3 \rangle = 0$ and
$\langle \Sigma^{}_2 \rangle \neq 0$, we get a more special
texture of $M^{}_\nu$ which can be diagonalized by the orthogonal
transformation
\begin{eqnarray}
V^{}_1 = \frac{1}{\sqrt{2}}\left( \matrix{ 1 & {\bf 0} & - 1 \cr
{\bf 0} & \sqrt{2} & {\bf 0} \cr 1 & {\bf 0} & 1}\right) \; .
\end{eqnarray}
The mass eigenvalues of $M^{}_\nu$ turn out to be $m^{}_1 =
|\delta m + \omega^{}_2|$, $m^{}_2 = |\delta m|$ and $m^{}_3 = |
\delta m - \omega^{}_2|$.  To be more explicit, we take $\delta m
>0$. Since $m^{}_1 < m^{}_2$ is required by current neutrino
oscillation data, we can obtain the normal neutrino mass hierarchy
by setting $\omega^{}_2 <0$. Then the ratio of two neutrino
mass-squared differences is given by $\Delta m^2_{21}/\Delta
m^2_{32} = (1-\alpha)/(1+\alpha)$ with $\alpha =
|\omega^{}_2|/2\delta m$. Taking $\Delta m^2_{21} \approx 8.0
\times 10^{-5} ~ {\rm eV}^2$ and $\Delta m^2_{32} \approx 2.5
\times 10^{-3} ~ {\rm eV}^2$ \cite{Strumia} as the typical inputs,
we obtain $\alpha \approx 0.94$, $|\omega^{}_2| \approx 0.035~{\rm
eV}$ and $\delta m \approx 0.019~{\rm eV}$.

The lepton flavor mixing matrix $V$ describes the mismatch between
the diagonalizations of $M^{}_l$ and $M^{}_\nu$ and is given by $V
= U^\dagger_l U^{}_1 V^{}_1$, where $U^{}_1$ and $V^{}_1$ have
generally been defined in Eq. (6). Note that the small deviation
of $U^{}_1$ from the unit matrix characterizes the unitarity
violation of $V$, while $V^{}_1$ is unitary and its expression has
been given in Eq. (21). In the approximation of $U^{}_1 \approx
{\bf 1}$, $V$ is just the tri-bimaximal mixing pattern
\cite{tribi} compatible with current experimental data:
\begin{eqnarray}
 V \approx U^\dagger_l U^{}_\nu = \left(\matrix{\frac{2}{\sqrt{6}} &
 \frac{1}{\sqrt{3}} & {\bf 0} \cr -\frac{1}{\sqrt{6}}\omega^2 &
 \frac{1}{\sqrt{3}}\omega^2 & -\frac{1}{\sqrt{2}}e^{-i\pi/6}
 \cr -\frac{1}{\sqrt{6}}\omega & \frac{1}{\sqrt{3}}\omega &
-\frac{1}{\sqrt{2}}e^{+i\pi/6}} \right) \; .
\end{eqnarray}
Thus this Type-II seesaw scenario is viable to interpret the
observed neutrino mass spectrum and neutrino mixing pattern.
Appreciable collider signatures can be achieved by adjusting the
ratio $m^{}_{\rm D}/m^{}_{\rm R}$, which is apparently independent
of the parameters responsible for the masses of light neutrinos
(i.e., $\delta m$ and $\omega^{}_2$), since the strength of
charged-current interactions of heavy Majorana neutrinos $N^{}_i$
is essentially described by $R \approx U^\dagger_l m^{}_{\rm
D}/m^{}_{\rm R}$. More discussions about the unitarity violation
of $V$ and possible collider signatures of $N^{}_i$ will be given
in section IV.

(2) Perturbations to $M^{}_{\rm D}$: $l^{}_{\rm L}$, $\chi$,
$\Sigma$ and $\phi$ are odd under the $Z^{}_2$ transformation,
while the other fields are even under the same transformation. In
this case, the Yukawa interaction $y^{}_{\Sigma}
\overline{l^{}_{\rm L}} i\sigma^{}_2 \Sigma l^c_{\rm L}$ is again
forbidden, so is the term $y^\prime_\nu \overline{l^{}_{\rm L}}
N^{}_{\rm R} \Phi$. However, one can resort to new scalar doublets
$\Phi^\prime$ --- their $A^{}_4\times U(1)^{}_{\rm X}$ charges are
the same as $\Phi$'s but their $Z^{}_2$ charge is opposite to
$\Phi$'s. Then the mass matrix $M^{}_{\rm D}$ takes the form
\begin{eqnarray}
M^{}_{\rm D} = m^{}_{\rm D} \left(\matrix{ 1 & {\bf 0} & \lambda
\cr {\bf 0} & 1 & {\bf 0} \cr \lambda & {\bf 0} & 1} \right) \; ,
\end{eqnarray}
where $\lambda = y^\prime_\nu \langle \Phi^\prime_2
\rangle/m^{}_{\rm D}$ and $\langle \Phi^\prime_1 \rangle = \langle
\Phi^\prime_3 \rangle = 0$, but the mass matrices $M^{}_{\rm L}$
and $M^{}_{\rm R}$ keep unchanged (i.e., $M^{}_{\rm L} = m^{}_{\rm
L} \cdot {\bf 1}$ and $M^{}_{\rm R} = m^{}_{\rm R} \cdot {\bf
1}$). Using the Type-II seesaw formula, we get
\begin{eqnarray}
M^{}_\nu = \delta m \cdot {\bf 1} - \frac{m^2_{\rm D}}{m^{}_{\rm R}}
\left( \matrix{\lambda^2 & {\bf 0} & 2\lambda \cr {\bf 0} & {\bf 0}
& {\bf 0} \cr 2\lambda & {\bf 0} & \lambda^2} \right) \; .
\end{eqnarray}
This effective neutrino mass matrix can also be diagonalized by
the orthogonal transformation given in Eq. (21). Its three
eigenvalues are found to be $m^{}_1 \approx |\delta m - 2\lambda
m^2_{\rm D}/m^{}_{\rm R}|$, $m^{}_2 = \delta m$ and $m^{}_3
\approx |\delta m + 2\lambda m^2_{\rm D}/m^{}_{\rm R}|$, where the
terms of ${\cal O}(\lambda^2)$ or smaller have been neglected.
Taking $\Delta m^2_{21} \approx 8.0 \times 10^{-5} ~ {\rm eV}^2$
and $\Delta m^2_{32} \approx 2.5 \times 10^{-3} ~ {\rm eV}^2$
\cite{Strumia} as the typical inputs, we obtain $\delta m \approx
0.019~{\rm eV}$ and $\lambda m^2_{\rm D}/m^{}_{\rm R} \approx
0.018~{\rm eV}$. Given $m^{}_{\rm R} \sim 100~{\rm GeV}$ and
$m^{}_{\rm D}/m^{}_{\rm R} \sim 0.1$ so as to make the heavy
Majorana neutrinos detectable at the LHC, the magnitude of
$\lambda$ turns out to be $\lambda \sim 10^{-11}$ in order to
generate the correct magnitude of light neutrino masses. Namely,
the smallness of $m^{}_i$ is attributed to the tiny perturbation
parameter $\delta m$ and the $U(1)^{}_{\rm X}$ symmetry breaking
parameter $\lambda$.

In this Type-II seesaw scenario, the lepton flavor mixing matrix
$V = U^\dagger_l U^{}_1 V^{}_1 \approx U^\dagger_l V^{}_1$ is the
same as that given in Eq. (22), where the small effects of
unitarity violation have been neglected. The strength of
charged-current interactions of heavy Majorana neutrinos can also
approximate to $R \approx U^{\dagger}_l m^{}_{\rm D}/m^{}_{\rm
R}$, because $\lambda$ is vanishingly small.

(3) Perturbations to $M^{}_{\rm R}$: $l^{}_{\rm L}$, $E^{}_{\rm
R}$, $E^{\prime}_{\rm R}$, $E^{\prime \prime}_{\rm R}$, $\Sigma$
and $\phi$ are odd under the $Z^{}_2$ transformation, while the
other fields are even under the same transformation. In this case,
the $Z^{}_2$-conserving term $y^{}_\chi \overline{N^{c}_{\rm R}}
N^{}_{\rm R} \chi$ exists. Then the right-handed Majorana neutrino
mass matrix reads
\begin{eqnarray}
M^{}_{\rm R} = m^{}_{\rm R} \left(\matrix{ 1 & {\bf 0} & \varrho
\cr {\bf 0} & 1 & {\bf 0} \cr \varrho & {\bf 0} & 1} \right) \; ,
\end{eqnarray}
where $\varrho = y^{}_\chi \langle \chi^{}_2 \rangle/m^{}_{\rm R}$
and $\langle \chi^{}_1 \rangle = \langle \chi^{}_3 \rangle = 0$,
but the mass matrices $M^{}_{\rm L}$ and $M^{}_{\rm D}$ keep
unchanged (i.e., $M^{}_{\rm L} = m^{}_{\rm L} \cdot {\bf 1}$ and
$M^{}_{\rm D} = m^{}_{\rm D} \cdot {\bf 1}$). From the Type-II
seesaw formula, we obtain
\begin{eqnarray}
M^{}_\nu = \delta m - \frac{m^2_{\rm D}}{m^{}_{\rm R}}
\frac{\varrho}{1-\varrho^2} \left( \matrix{\varrho & {\bf 0} & -1
\cr {\bf 0} & {\bf 0} & {\bf 0} \cr -1 & {\bf 0} & \varrho} \right)
\; .
\end{eqnarray}
The orthogonal transformation in Eq. (21) can also be used to
diagonalize the effective neutrino mass matrix in Eq. (26). After
a straightforward calculation, we get $m^{}_1 \approx |\delta m +
\varrho m^2_{\rm D}/m^{}_{\rm R}|$, $m^{}_2 \approx \delta m$ and
$m^{}_3 \approx |\delta m - \varrho m^2_{\rm D}/m^{}_{\rm R}|$,
where the terms of ${\cal O}(\varrho^2)$ or smaller have been
omitted. Then $\delta m \approx 0.019~{\rm eV}$ and
$|\varrho|m^2_{\rm D}/m^{}_{\rm R} \approx 0.035~{\rm eV}$ are
obtained from the typical inputs $\Delta m^2_{21} \approx 8.0
\times 10^{-5} ~ {\rm eV}^2$ and $\Delta m^2_{32} \approx 2.5
\times 10^{-3} ~ {\rm eV}^2$ \cite{Strumia}. Given $m^{}_{\rm R}
\sim 100~{\rm GeV}$ and $m^{}_{\rm D}/m^{}_{\rm R} \sim 0.1$ to
make the heavy Majorana neutrinos detectable at the LHC, the sign
and magnitude of $\varrho$ are required to be $\varrho < 0$ and
$|\varrho| \sim 10^{-11}$ by current neutrino oscillation data. In
this scenario, the lepton flavor mixing matrix $V = U^\dagger_l
U^{}_1 V^{}_1 \approx U^\dagger_l V^{}_1$ is the same as that
given in Eq. (22), where the small effects of unitarity violation
have been neglected. The strength of charged-current interactions
of heavy Majorana neutrinos can also approximate to $R \approx
U^{\dagger}_l m^{}_{\rm D}/m^{}_{\rm R}$, due to the smallness of
$\varrho$.

The scenarios given above illustrate three simple ways to deform
the complete cancellation between $M^{}_{\rm L}$ and $M^{}_{\rm D}
M^{-1}_{\rm R} M^T_{\rm D}$ terms such that tiny neutrino masses
can be generated through the Type-II seesaw formula. A general
approach should include the perturbations to $M^{}_{\rm L}$,
$M^{}_{\rm D}$ and $M^{}_{\rm R}$ together. Let us denote
$M^{}_{\rm L, D, R}$ as a sum of the ``symmetry" term and the
``perturbation" term: $M^{}_{\rm L, D, R} = \tilde{M}^{}_{\rm L,
D, R} + \delta M^{}_{\rm L, D, R}$. The residue of the incomplete
cancellation between $\tilde{M}^{}_{\rm L}$ and $\tilde{M}^{}_{\rm
D} \tilde{M}^{-1}_{\rm R} \tilde{M}^T_{\rm D}$ terms is denoted by
$\delta M$ (i.e., $\delta M = \tilde{M}^{}_{\rm L} -
\tilde{M}^{}_{\rm D} \tilde{M}^{-1}_{\rm R} \tilde{M}^T_{\rm D}$).
Then the Type-II seesaw formula $M^{}_\nu \approx M^{}_{\rm L} -
M^{}_{\rm D} M^{-1}_{\rm R} M^T_{\rm D}$ can be re-expressed as
\begin{eqnarray}
M^{}_\nu \approx \delta M + \delta M^{}_{\rm L} +
\tilde{M}^{}_{\rm D} \tilde{M}^{-1}_{\rm R} \delta M^{}_{\rm R}
\tilde{M}^{-1}_{\rm R} \tilde{M}^T_{\rm D}- \tilde{M}^{}_{\rm D}
\tilde{M}^{-1}_{\rm R} (\delta M^{}_{\rm D})^T - \delta M^{}_{\rm
D} \tilde{M}^{-1}_{\rm R} \tilde{M}^T_{\rm D} \;
\end{eqnarray}
to the first order of $\delta M^{}_{\rm L, D, R}$. It is easy to
see that Eqs. (20), (24) and (26) are just the special cases of
Eq. (27).

We have shown that it is possible to achieve a phenomenological
compromise between tiny neutrino masses and accessible collider
signatures in the Type-II seesaw scenarios with spontaneous and
explicit breaking of the $A^{}_4 \times U(1)^{}_{\rm X}$ flavor
symmetry. Proper $A^{}_4$ symmetry breaking is also necessary in the
quark sector to account for the observed quark mass spectra and
flavor mixing parameters, as discussed in Ref. \cite{He}. Note that
radiative corrections to the light neutrino masses may be very large
due to the largeness of Yukawa interactions in a certain Type-I or
Type-II seesaw model, but some detailed calculations have shown that
these corrections are vanishing (or vanishingly small) in the limit
of degenerate (or nearly degenerate) heavy Majorana neutrino masses
\cite{PU}. This is just the case for three simple Type-II seesaw
scenarios discussed above. On the other hand, the seesaw threshold
effects are also negligible in our examples because of the (near)
mass degeneracy of three heavy Majorana neutrinos.

\section{Unitarity Violation and Collider Signatures}

Now we proceed to discuss the unitarity violation and collider
signatures in the Type-II seesaw model. The non-unitarity of the
lepton flavor mixing matrix $V$ is actually a common feature of
the seesaw models, as one can easily see from $VV^\dagger = {\bf
1} - RR^\dagger \neq {\bf 1}$ in Eq. (4a). Taking account of Eqs.
(3), (5), (6) and (9), we may express $V$ and $R$ as $V =
U^\dagger_l U^{}_1 V^{}_1$ and $R = U^\dagger_l B V^{}_2$, where
$U^{}_l$ is the unitary matrix defined to diagonalize the
Hermitian matrix $M^{}_l M^\dagger_l$ with $M^{}_l$ being the
charge-lepton mass matrix. The $3\times 3$ matrices $U^{}_1$, $B$,
$V^{}_1$ and $V^{}_2$ can in principle be determined by the
neutrino mass matrices $M^{}_{\rm L}$, $M^{}_{\rm D}$ and
$M^{}_{\rm R}$, and thus $V$ and $R$ should be calculable. In
practice, one may resort to a recursive expansion of $M^{}_\nu$ in
powers of $M^{}_{\rm D} M^{-1}_{\rm R}$ by taking the reasonable
assumptions $C=-B^\dagger$, $U^{}_1 = \sqrt{{\bf 1} -B B^\dagger}$
and $U^{}_2 = \sqrt{{\bf 1} -B^\dagger B}$ \cite{Grimus}. Then
$U^{}_1 \approx {\bf 1} - BB^\dagger/2$ and $B \approx U^\dagger_l
M^{}_{\rm D} M^{-1}_{\rm R}$ are two good approximations, from
which
\begin{eqnarray}
V \approx U^\dagger_l \left[ {\bf 1} - \frac{1}{2} U^\dagger_l
M^{}_{\rm D} M^{-1}_{\rm R} (M^{-1}_{\rm R} M^T_{\rm D})^* U^{}_l
\right] V^{}_1 \;
\end{eqnarray}
can be obtained. For simplicity, let us define $\xi \equiv
{U^\dagger_l}^2 M^{}_{\rm D} M^{-1}_{\rm R} (M^{-1}_{\rm R}
M^T_{\rm D})^* {U^{}_l}^2$. Note that the Hermitian matrix $\xi$
is suppressed by two powers of $M^{}_{\rm D} M^{-1}_{\rm R}$.
Hence $V \approx U^\dagger_l V^{}_1$ is unitary in the
leading-order approximation \cite{xingzhou}. To a better degree of
accuracy, we have $V \approx ({\bf 1} - \xi/2)U^\dagger_l V^{}_1$
and $V V^\dagger \approx {\bf 1} - \xi$. Then we arrive at $\xi
\approx R R^\dagger$. Note also that $\xi$ is in general complex
and may give rise to some additional CP-violating effects in
neutrino oscillations \cite{CPV}. In the framework of two-flavor
oscillations, where the non-trivial CP-violating phase of
$U^\dagger_l V^{}_1$ is negligible, it remains possible to get a
CP-violating asymmetry between the probabilities of $\nu^{}_\alpha
\to \nu^{}_\beta$ and $\bar{\nu}^{}_\alpha \to \bar{\nu}^{}_\beta$
transitions:
\begin{eqnarray}
\frac{P(\nu^{}_\alpha \to \nu^{}_\beta) - P(\bar{\nu}^{}_\alpha
\to \bar{\nu}^{}_\beta)}{P(\nu^{}_\alpha \to \nu^{}_\beta) +
P(\bar{\nu}^{}_\alpha \to \bar{\nu}^{}_\beta)} \propto \left|
\xi^{}_{\alpha \beta}\right| \sin \delta^{}_{\alpha \beta} \;
\end{eqnarray}
with $\delta^{}_{\alpha \beta} \equiv \arg(\xi^{}_{\alpha \beta})$
for $\alpha, \beta = e, \mu, \tau$ \cite{CPV}. When the specific
Type-II seesaw scenarios proposed in section III are taken into
account, we find $\xi \approx R R^\dagger \approx m^2_{\rm
D}/m^2_{\rm R} \cdot {\bf 1}$ and thus $\delta^{}_{\alpha\beta}
\approx 0$. This result shows that there is almost no extra CP
violation induced by the unitarity violation of $V$ in our special
examples. Nevertheless, the diagonal elements of $\xi$ can be as
large as ${\cal O}(10^{-2})$ for $m^{}_{\rm D}/m^{}_{\rm R} \sim
{\cal O}(10^{-1})$, implying that the deviation of $V$ from
unitarity can actually reach the percent level. It is worth
emphasizing that such a model-dependent argument has no conflict
with the model-independent bound on $VV^\dagger$ or equivalently
on $\xi$. A global analysis of current neutrino oscillation data
and precision electroweak data (e.g., on the invisible width of
the $Z^0$ boson, universality tests and rare lepton decays) has
yielded quite strong constraints on the unitarity of $V$ and its
possible violation \cite{antusch}. Translating the numerical
results of Refs. \cite{CPV} and \cite{antusch} into the
restriction on $\xi$ in our language, we obtain
\begin{eqnarray}
\left|\xi\right| = \left(\matrix{|\xi^{}_{ee}| < 1.1 \cdot 10^{-2}
& |\xi^{}_{e\mu}| < 7.0 \cdot 10^{-5} & |\xi^{}_{e\tau}| < 1.6
\cdot 10^{-2} \cr |\xi^{}_{\mu e}| < 7.0 \cdot 10^{-5} &
|\xi^{}_{\mu\mu}| < 1.0 \cdot 10^{-2} & |\xi^{}_{\mu \tau}| < 1.0
\cdot 10^{-2} \cr |\xi^{}_{\tau e}| < 1.6 \cdot 10^{-2} &
|\xi^{}_{\tau \mu}| < 1.0 \cdot 10^{-2} & |\xi^{}_{\tau\tau}| <
1.0 \cdot 10^{-2} \cr}\right) \;
\end{eqnarray}
at the $90\%$ confidence level. It is clear that the effects of
unitarity violation can saturate the experimental upper bounds in
our Type-II seesaw scenarios, only if $m^{}_{\rm D}/m^{}_{\rm R}
\sim 0.1$ is taken. The latter may lead to appreciable collider
signatures of lepton number violation induced by the heavy
Majorana neutrinos and doubly-charged scalars.

A direct test of the seesaw mechanism requires the unambiguous
observation of heavy Majorana neutrinos. The clearest signature
induced by $N^{}_i$ should be the lepton-number-violating process
$pp \to W^\pm \to \mu^\pm N \to \mu^\pm \mu^\pm jj$ at the LHC
\cite{Han,LHC}. For the doubly-charged scalars existing in the
Type-II seesaw model, one may concentrate on either the single
production $pp \to W^\pm W^\pm \to \Delta^{\pm \pm}$ \cite{single}
or the pair production in the Drell-Yan process $q \bar{q} \to
\gamma^*/ Z^* \to \Delta^{\pm \pm} \Delta^{\mp \mp}$ \cite{pair}
and the subsequent decays $\Delta^{\pm \pm} \to W^{\pm} W^{\pm}$
or $\Delta^{\pm \pm} \to l^{\pm} l^{\pm}$. Some remarks are in
order:
\begin{itemize}
\item The lepton-number-violating processes include both $pp \to
W^\pm W^\pm \to \mu^\pm \mu^\pm jj$ and $pp \to W^\pm \to \mu^\pm
N \to \mu^\pm \mu^\pm jj$ modes. The latter can be resonantly
enhanced due to the on-shell production of heavy Majorana
neutrinos. Given $M^{}_i \sim 100~{\rm GeV}$ for example, one may
follow the analysis of Ref. \cite{Han} to show that it is possible
to probe $\xi^{}_{\mu \mu}$ of ${\cal O}(10^{-4})$ at the
$2\sigma$ level by means of the LHC with an integrated luminosity
$100 ~{\rm fb^{-1}}$. Even though the background might be more
complicated than naively expected \cite{LHC}, we feel that the
discovery of heavy Majorana neutrinos with $M^{}_i \sim {\cal
O}(10^2)~{\rm GeV}$ to ${\cal O}(1)~{\rm TeV}$ and $\xi^{}_{\mu
\mu} \sim {\cal O}(10^{-3})$ to ${\cal O}(10^{-2})$ is still
possible.

\item Because light neutrino masses arise from the significant
cancellation between $M^{}_{\rm L}$ and $M^{}_{\rm D} M^{-1}_{\rm R}
M^T_{\rm D}$ terms in our Type-II seesaw scenarios, one can notice
that $m^{}_{\rm L}$ is much larger than $m^{}_i$. Taking $m^{}_{\rm
R} \sim 100 ~{\rm GeV}$ and $m^{}_{\rm D}/m^{}_{\rm R} \sim 0.1$ for
example, we obtain $m^{}_{\rm L} \approx m^2_{\rm D}/m^{}_{\rm R}
\sim 1$ GeV as a consequence of cancellation. The implication of
$m^{}_{\rm L} = y^{}_\Delta \langle \Delta \rangle \sim 1~{\rm GeV}$
is rather clear: even if the vev of the Higgs triplet reaches the
experimental upper bound $\langle \Delta \rangle \lesssim 1~{\rm
GeV}$, one can get a large Yukawa coupling $y^{}_\Delta \sim {\cal
O}(1)$. The single production rate of $W^{\pm} W^\pm \to \Delta^{\pm
\pm}$ is proportional to $(\langle \Delta \rangle/v)^2 \sim
10^{-4}$, so this process is too small to be observed at the LHC. In
Ref. \cite{pair}, it has been advocated that signatures of the
doubly-charged scalars can be observed at the LHC via the pair
production channel and the $l^\pm l^\pm$ decay mode with a branching
fraction $\sim 50\%$ up to the mass range of $800~{\rm GeV}$ to
$1~{\rm TeV}$. This conclusion is applicable to our model, but the
choice of $y^{}_\Delta \sim {\cal O}(1)$ and $\langle \Delta \rangle
\sim 1~{\rm GeV}$ will extend the above mass range for the
doubly-charged scalars. As the total decay rate is enlarged,
however, the $\Delta^{\pm \pm}$ particles cannot be the long-lived
doubly-charged scalars which have been looked for at the Tevatron.
\end{itemize}
Of course, it is also possible to search for the
lepton-number-violating signatures at the future International
Linear Collider (ILC) via the processes $e^+ e^- \to W^\pm/Z^* \to
\nu N$ for the heavy Majorana neutrinos and $e^+ e^- \to
\gamma^*/Z^* \to \Delta^{\pm \pm} \Delta^{\mp \mp}$ for the
doubly-charged scalars.

\section{Concluding remarks}

The main concern of this work is the experimental testability of
the seesaw mechanism in the era of LHC and (or) ILC. We have
presented the most general proof of a no-go theorem, which forbids
the tree-level generation of light Majorana neutrino masses if the
condition $M^{}_{\rm L} = M^{}_{\rm D} M^{-1}_{\rm R} M^T_{\rm D}$
is satisfied in the Type-II seesaw model. Furthermore, we have
shown that a compromise between tiny neutrino masses and
appreciable collider signatures can be achieved by allowing for a
significant but incomplete cancellation between $M^{}_{\rm L}$ and
$M^{}_{\rm D} M^{-1}_{\rm R} M^T_{\rm D}$ terms. In other words,
observable effects of lepton number violation may be induced by
the heavy Majorana neutrinos and doubly-charged scalars at the TeV
scale because both $M^{}_{\rm L}$ and $M^{}_{\rm D} M^{-1}_{\rm R}
M^T_{\rm D}$ terms are not strongly suppressed, but their
difference is tiny and responsible for the tiny masses of three
light Majorana neutrinos. We have proposed three simple but viable
Type-II seesaw scenarios, in which the $A^{}_4 \times U(1)^{}_{\rm
X}$ flavor symmetry is taken into account, to illustrate our main
ideas.

It is worth highlighting that the non-unitarity of the lepton
flavor mixing matrix $V$, which describes the strength of
charged-current interactions of light Majorana neutrinos, is an
intrinsic feature of the seesaw models. The CP-conserving and
CP-violating effects of this unitarity violation can be measured
or constrained in the future long-baseline neutrino oscillation
experiments.

It is also worth remarking the interesting correlation between $V$
and $R$, the $3\times 3$ rotation matrix which characterizes the
strength of charged-current interactions of heavy Majorana
neutrinos. As a result of $VV^\dagger = {\bf 1} - R R^\dagger$ in
both Type-I and Type-II seesaw models, larger magnitudes of the
elements of $R$ lead to larger deviations of $V$ from unitarity
(or vice versa). In this sense, testing the unitarity of $V$ in
neutrino oscillations and searching for heavy Majorana neutrinos
at hadron or $e^+e^-$ colliders are the two faces of one coin:
they can be complementary to each other, both qualitatively and
quantitatively, to understand the properties of light and heavy
Majorana neutrinos.

Although the Type-II seesaw scenarios proposed in this paper are
far from perfect, they may serve as a phenomenological example to
illustrate possible ways for model building. But much more efforts
are certainly needed to study neutrino physics at the TeV scale.
For instance, one may question whether a compromise can still be
achieved between tiny neutrino masses and appreciable collider
signatures, when a successful realization of the TeV-scale
leptogenesis is simultaneously required. We shall address
ourselves to such difficult but interesting problems elsewhere.

\begin{acknowledgments}
This work was supported in part by the National Natural Science
Foundation of China.
\end{acknowledgments}

\appendix

\section{The Scalar Potential}

In this appendix, we list the complete scalar potential in the
type-II seesaw scenarios proposed in section III. For simplicity,
only the first scenario is considered, and the other two cases can
be discussed in a similar way. The $SU(2)^{}_{\rm L} \times
U(1)^{}_{\rm Y} \otimes A^{}_4$ invariant and renormalizable terms
with the discrete $Z^{}_2$ symmetry can in general be written as
\begin{eqnarray}
V(\Phi) &=& \mu^2_{\Phi} \left(\Phi^\dagger
\Phi\right)^{}_{\underline{1}} + \lambda^\Phi_1 \left(\Phi^\dagger
\Phi\right)^{}_{\underline{1}} \left(\Phi^\dagger
\Phi\right)^{}_{\underline{1}} + \lambda^\Phi_2 \left(\Phi^\dagger
\Phi\right)^{}_{\underline{1}^\prime} \left(\Phi^\dagger
\Phi\right)^{}_{\underline{1}^{\prime \prime}} \nonumber \\
&& + \lambda^\Phi_3 \left(\Phi^\dagger
\Phi\right)^{}_{\underline{3_{\rm s}}} \left(\Phi^\dagger
\Phi\right)^{}_{\underline{3_{\rm s}}} + \lambda^\Phi_4
\left(\Phi^\dagger \Phi\right)^{}_{\underline{3_{\rm a}}}
\left(\Phi^\dagger \Phi\right)^{}_{\underline{3_{\rm a}}} \nonumber
\\
&& + i\lambda^\Phi_5 \left(\Phi^\dagger
\Phi\right)^{}_{\underline{3_{\rm s}}} \left(\Phi^\dagger
\Phi\right)^{}_{\underline{3_{\rm a}}} \; ,\\
V(\chi) &=& \mu^2_\chi \left(\chi^\dagger
\chi\right)^{}_{\underline{1}} + \lambda^\chi_1 \left(\chi^\dagger
\chi\right)^{}_{\underline{1}} \left(\chi^\dagger
\chi\right)^{}_{\underline{1}}  + \lambda^\chi_2 \left(\chi^\dagger
\chi\right)^{}_{\underline{1}^\prime} \left(\chi^\dagger
\chi\right)^{}_{\underline{1}^{\prime \prime}}  \nonumber
\\
&& + \lambda^\chi_3 \left(\chi^\dagger
\chi\right)^{}_{\underline{3_{\rm s}}} \left(\chi^\dagger
\chi\right)^{}_{\underline{3_{\rm s}}} + \lambda^\chi_4
\left(\chi^\dagger \chi\right)^{}_{\underline{3_{\rm a}}}
\left(\chi^\dagger \chi\right)^{}_{\underline{3_{\rm a}}} \nonumber
\\
&& + i \lambda^\chi_5 \left(\chi^\dagger
\chi\right)^{}_{\underline{3_{\rm s}}}
\left(\chi^\dagger \chi\right)^{}_{\underline{3_{\rm a}}}\;, \\
V(\phi) &=& \mu^2_\phi \left(\phi^\dagger
\phi\right)^{}_{\underline{1}} + \lambda^\phi \left(\phi^\dagger
\phi\right)^2_{\underline{1}} \; , \\
V(\Delta) &=& \mu^2_\Delta {\rm Tr}\left(\Delta^\dagger
\Delta\right)^{}_{\underline{1}} + \lambda^\Delta_1 {\rm
Tr}\left(\Delta^\dagger \Delta\right)^{}_{\underline{1}} {\rm
Tr}\left(\Delta^\dagger \Delta\right)^{}_{\underline{1}} \nonumber \\
&& + \lambda^\Delta_2 {\rm Tr}\left[\left(\Delta^\dagger
\Delta\right)^{}_{\underline{1}}
\left(\Delta^\dagger \Delta\right)^{}_{\underline{1}}\right] \; ,\\
V(\Sigma) &=& \mu^2_\Sigma {\rm Tr} \left(\Sigma^\dagger
\Sigma\right)^{}_{\underline{1}} + \lambda^\Sigma_1 {\rm
Tr}\left(\Sigma^\dagger \Sigma\right)^{}_{\underline{1}} {\rm
Tr}\left(\Sigma^\dagger \Sigma\right)^{}_{\underline{1}} +
\lambda^\Sigma_2 {\rm Tr}\left(\Sigma^\dagger
\Sigma\right)^{}_{\underline{1}^\prime} {\rm Tr}\left(\Sigma^\dagger
\Sigma\right)^{}_{\underline{1}^{\prime \prime}} \nonumber \\
&& + \lambda^\Sigma_3 {\rm Tr}\left(\Sigma^\dagger
\Sigma\right)^{}_{\underline{3_{\rm s}}} {\rm
Tr}\left(\Sigma^\dagger \Sigma\right)^{}_{\underline{3_{\rm s}}} +
\lambda^\Sigma_4 {\rm Tr}\left(\Sigma^\dagger
\Sigma\right)^{}_{\underline{3_{\rm a}}} {\rm
Tr}\left(\Sigma^\dagger \Sigma\right)^{}_{\underline{3_{\rm a}}}
\nonumber
\\
&& + i\lambda^\Sigma_5 {\rm Tr}\left(\Sigma^\dagger
\Sigma\right)^{}_{\underline{3_{\rm s}}} {\rm
Tr}\left(\Sigma^\dagger
\Sigma\right)^{}_{\underline{3_{\rm a}}} + \lambda^\Sigma_6 {\rm Tr}\left[
\left(\Sigma^\dagger \Sigma\right)^{}_{\underline{1}}
\left(\Sigma^\dagger \Sigma\right)^{}_{\underline{1}}\right]
\nonumber \\
&& + \lambda^\Sigma_7 {\rm Tr}\left[ \left(\Sigma^\dagger
\Sigma\right)^{}_{\underline{1}^\prime} \left(\Sigma^\dagger
\Sigma\right)^{}_{\underline{1}^{\prime \prime}}\right] +
\lambda^\Sigma_8 {\rm Tr}\left[
\left(\Sigma^\dagger \Sigma\right)^{}_{\underline{3_{\rm s}}}
\left(\Sigma^\dagger \Sigma\right)^{}_{\underline{3_{\rm s}}}\right]
\nonumber \\
&& + \lambda^\Sigma_9 {\rm Tr}\left[ \left(\Sigma^\dagger
\Sigma\right)^{}_{\underline{3_{\rm a}}} \left(\Sigma^\dagger
\Sigma\right)^{}_{\underline{3_{\rm a}}}\right] +
i\lambda^\Sigma_{10} {\rm Tr}\left[
\left(\Sigma^\dagger \Sigma\right)^{}_{\underline{3_{\rm s}}}
\left(\Sigma^\dagger \Sigma\right)^{}_{\underline{3_{\rm a}}}\right]\; ,\\
V(\Phi, \chi) &=& \lambda^{\Phi \chi}_1 \left(\Phi^\dagger
\Phi\right)^{}_{\underline{1}} \left(\chi^\dagger
\chi\right)^{}_{\underline{1}} + \lambda^{\Phi \chi}_2
\left(\Phi^\dagger \Phi\right)^{}_{\underline{1}^\prime}
\left(\chi^\dagger \chi\right)^{}_{\underline{1}^{\prime \prime}} +
\lambda^{\Phi \chi}_3 \left(\Phi^\dagger
\Phi\right)^{}_{\underline{1}^{\prime \prime}} \left(\chi^\dagger
\chi\right)^{}_{\underline{1}^{\prime}} \nonumber \\
&& + \lambda^{\Phi \chi}_4 \left(\Phi^\dagger
\Phi\right)^{}_{\underline{3_{\rm s}}} \left(\chi^\dagger
\chi\right)^{}_{\underline{3_{\rm s}}} + \lambda^{\Phi \chi}_5
\left(\Phi^\dagger \Phi\right)^{}_{\underline{3_{\rm a}}}
\left(\chi^\dagger \chi\right)^{}_{\underline{3_{\rm a}}} \nonumber
\\
&& + i\lambda^{\Phi \chi}_6 \left(\Phi^\dagger
\Phi\right)^{}_{\underline{3_{\rm s}}} \left(\chi^\dagger
\chi\right)^{}_{\underline{3_{\rm a}}} + i\lambda^{\Phi \chi}_7
\left(\Phi^\dagger \Phi\right)^{}_{\underline{3_{\rm a}}}
\left(\chi^\dagger \chi\right)^{}_{\underline{3_{\rm s}}} \; , \\
V(\Phi, \phi) &=& \left[ \lambda^{\Phi \phi}_1 \left(\Phi^\dagger
\phi\right)^{}_{\underline{3}} \left(\Phi^\dagger
\phi\right)^{}_{\underline{3}} + {\rm h.c.} \right] + \lambda^{\Phi
\phi}_2 \left(\Phi^\dagger \phi\right)^{}_{\underline{3}}
\left(\phi^\dagger \Phi\right)^{}_{\underline{3}} \nonumber \\
&& + \lambda^{\Phi \phi}_3 \left(\Phi^\dagger
\Phi\right)^{}_{\underline{1}} \left(\phi^\dagger
\phi\right)^{}_{\underline{1}} \; , \\
V(\Phi, \Delta) &=& \lambda^{\Phi \Delta}_1 \left(\Phi^\dagger
\Phi\right)^{}_{\underline{1}} {\rm Tr} \left(\Delta^\dagger
\Delta\right)^{}_{\underline{1}} + \lambda^{\Phi \Delta}_2 \Phi^\dagger_{\underline{3}}
\left[\Delta, \Delta^\dagger \right]^{}_{\underline{1}} \Phi^{}_{\underline{3}}\; , \\
V(\Phi, \Sigma) &=& \lambda^{\Phi \Sigma}_1 \left(\Phi^\dagger
\Phi\right)^{}_{\underline{1}} {\rm Tr} \left(\Sigma^\dagger
\Sigma\right)^{}_{\underline{1}} + \lambda^{\Phi \Sigma}_2
\Phi^\dagger \left[\Sigma, \Sigma^\dagger \right] \Phi
+ \left[\lambda^{\Phi \Sigma}_3 \left(\Phi^T \Phi\right)^{}_{}
\Sigma^{}_{}+ {\rm h.c.} \right]\; , \\
V(\chi, \phi) &=& \lambda^{\chi \phi}_{1} \left(\chi^\dagger
\chi\right)^{}_{\underline{1}} \left(\phi^\dagger
\phi\right)^{}_{\underline{1}} \; , \\
V(\chi, \Delta) &=& \lambda^{\chi \Delta}_{1} \left(\chi^\dagger
\chi\right)^{}_{\underline{1}} {\rm Tr} \left(\Delta^\dagger
\Delta\right)^{}_{\underline{1}} \; , \\
V(\chi, \Sigma) &=& \lambda^{\chi \Sigma}_{1} \left(\chi^\dagger
\chi\right)^{}_{\underline{1}} {\rm Tr} \left(\Sigma^\dagger
\Sigma\right)^{}_{\underline{1}} \; , \\
V(\phi, \Delta) &=& \lambda^{\phi \Delta}_1 \phi^\dagger \phi {\rm
Tr}\left(\Delta^\dagger \Delta\right) + \lambda^{\phi \Delta}_2
\phi^\dagger \left[\Delta, \Delta^\dagger\right] \phi + \left[
\lambda^{\phi \Delta}_3
\phi^T \phi \Delta^\dagger + {\rm h.c.} \right] \; , \\
V(\phi, \Sigma) &=& \lambda^{\phi \Sigma}_1 \phi^\dagger \phi {\rm
Tr}\left(\Sigma^\dagger \Sigma\right) +
\lambda^{\phi \Sigma}_2 \phi^\dagger \left[\Sigma, \Sigma^\dagger\right] \phi \; ,\\
V(\Delta, \Sigma) &=& \lambda^{\Delta \Sigma}_1 {\rm Tr}
\left(\Delta^\dagger \Delta\right)^{}_{\underline{1}} {\rm Tr}
\left(\Sigma^\dagger \Sigma\right)^{}_{\underline{1}} +
\lambda^{\Delta \Sigma}_2 {\rm Tr} \left[\left(\Delta^\dagger
\Delta\right)^{}_{\underline{1}} \left(\Sigma^\dagger
\Sigma\right)^{}_{\underline{1}} \right] \nonumber \\
&& + \lambda^{\Delta \Sigma}_3 {\rm Tr} \left[\left(\Delta^\dagger
\Sigma\right)^{}_{\underline{3}} \left(\Delta
\Sigma^\dagger\right)^{}_{\underline{3}} \right]\; ,\\
V(\Phi, \phi, \chi) &=& \lambda^{\Phi \phi \chi} (\phi^\dagger
\Phi)^{}_{\underline{3}} \chi^{}_{\underline{3}} + {\rm h.c.} \; .
\end{eqnarray}
Note that the above scalar potential also respects the $U(1)^{}_{\rm
X}$ symmetry except for the terms of $V(\Phi, \phi)$ in the square
bracket in Eq. (A7), which explicitly breaks $U(1)^{}_{\rm X}$ to
$Z^{}_2$.

\end{document}